\documentclass[conference]{IEEEtran}
\IEEEoverridecommandlockouts
\usepackage{cite}
\usepackage{amsmath,amssymb,amsfonts}
\usepackage{algorithmic}
\usepackage{graphicx}
\usepackage{textcomp}
\usepackage{xcolor}
\usepackage{float}
\usepackage{hyperref}

\usepackage{tikzit}

\tikzstyle{blueRec}=[fill={rgb,255: red,160; green,203; blue,255}, draw=none, shape=rectangle, tikzit shape=rectangle, tikzit fill={rgb,255: red,160; green,203; blue,255}]
\tikzstyle{blueLargeRec}=[fill={rgb,255: red,188; green,226; blue,255}, draw=black, shape=rectangle, minimum width=84pt, minimum height=16pt, line width=1.pt]
\tikzstyle{greenLargeRec}=[fill={rgb,255: red,210; green,255; blue,199}, draw=black, shape=rectangle, minimum width=84pt, minimum height=16pt, line width=1.pt]
\tikzstyle{redLargeRec}=[fill={rgb,255: red,255; green,191; blue,191}, draw=black, shape=rectangle, minimum width=84pt, minimum height=16pt, line width=1.pt]
\tikzstyle{yellowLargRec}=[fill={rgb,255: red,255; green,250; blue,189}, draw=black, shape=rectangle, minimum width=84pt, minimum height=16pt, line width=1.pt]
\tikzstyle{etc}=[fill=black, draw=black, shape=circle, minimum width=2pt, minimum height=2pt]

\tikzstyle{fillBlue}=[-, fill={rgb,255: red,167; green,225; blue,255}]
\tikzstyle{fillGreen}=[-, fill={rgb,255: red,201; green,255; blue,189}]
\tikzstyle{fillYellow}=[-, fill={rgb,255: red,255; green,253; blue,183}]
\tikzstyle{myEdgeArrows}=[->, line width=1.pt]
\tikzstyle{myEdge}=[-, line width=1.pt]
\tikzstyle{new edge style 0}=[-, fill=white]
\tikzstyle{myEdge2}=[<-, line width=1.]
\tikzstyle{dataflow}=[draw={rgb,255: red,191; green,0; blue,64}, ->, line width=5pt]
\tikzstyle{sync}=[-, draw=red, line width=2pt]

\def\BibTeX{{\rm B\kern-.05em{\sc i\kern-.025em b}\kern-.08em
    T\kern-.1667em\lower.7ex\hbox{E}\kern-.125emX}}

\usepackage[normalem]{ulem}
\usepackage{color}

\begin{document}

\title{Understanding the Impact of Synchronous, Asynchronous, and Hybrid In-Situ Techniques in Computational Fluid Dynamics Applications
}

\author{
\IEEEauthorblockN{Yi Ju \IEEEauthorrefmark{1}, Adalberto Perez \IEEEauthorrefmark{2}, Stefano Markidis \IEEEauthorrefmark{2}, Philipp Schlatter \IEEEauthorrefmark{2}, Erwin Laure \IEEEauthorrefmark{1}}
\IEEEauthorblockA{\IEEEauthorrefmark{1}\textit{Max Planck Computing and Data Facility}
 \{yi.ju, erwin.laure\}@mpcdf.mpg.de} \IEEEauthorblockA{\IEEEauthorrefmark{2}\textit{KTH Royal Institute of Technology}
 \{adperez, markidis, pschlatt\}@kth.se} 
}


\maketitle

\begin{abstract}

High-Performance Computing (HPC) systems provide input/output (IO) performance growing relatively slowly compared to peak computational performance and have limited storage capacity. Computational Fluid Dynamics (CFD) applications aiming to leverage the full power of Exascale HPC systems, such as the solver Nek5000, will generate massive data for further processing. These data need to be efficiently stored via the IO subsystem. However, limited IO performance and storage capacity may result in performance, and thus scientific discovery, bottlenecks. In comparison to traditional post-processing methods, in-situ techniques can reduce or avoid writing and reading the data through the IO subsystem, promising to be a solution to these problems. In this paper, we study the performance and resource usage of three in-situ use cases: data compression, image generation, and uncertainty quantification. We furthermore analyze three approaches when these in-situ tasks and the simulation are executed synchronously, asynchronously, or in a hybrid manner. In-situ compression can be used to reduce the IO time and storage requirements while maintaining data accuracy. Furthermore, in-situ visualization and analysis can save Terabytes of data from being routed through the IO subsystem to storage. However, the overall efficiency is crucially dependent on the characteristics of both, the in-situ task and the simulation. In some cases, the overhead introduced by the in-situ tasks can be substantial. Therefore, it is essential to choose the proper in-situ approach, synchronous, asynchronous, or hybrid, to minimize overhead and maximize the benefits of concurrent execution.  
\end{abstract}

\begin{IEEEkeywords}
CFD, in-situ, HPC
\end{IEEEkeywords}

\section{Introduction}
Computational fluid dynamics (CFD) is a branch of engineering physics that analyzes fluid flow problems using numerical methods. It is used to solve a wide range of problems in both research and industry. Examples include nuclear reactor flow analysis \cite{merzari2020toward}, biological flows e.g. food and drug administration (FDA) nozzle benchmark \cite{sanchez2020simulation}, and flow simulations around a wing for modern civil aircraft design \cite{hosseini2016direct}. Analyzing the flow around objects in realistic problems entails the analysis of turbulence, which is a flow regime characterized by complex, non-linear, and seemingly random fluid motions on multi-scales. 
Because the energy in turbulence is dissipated through viscosity at small scales, the discrete domains used to solve such problems must be large enough to capture the large scale motions and fine enough to capture the smallest ones. Direct Numerical Simulations (DNS) allow for full flow resolution but require domains with the order of hundreds of million grid points and hundreds of thousand time steps~\cite{res}.

In this paper, we focus on {\it Nek5000}\cite{NEK}, a spectral element method-based code with excellent scalability \cite{offermans} partially due to the weak element coupling and its ``matrix-free" formulation. This enables the solution of large problems without the need to explicitly construct any matrix operator, which would have restrictive sizes due to the large number of grid points in turbulence simulations. The code stores and processes information on a ``local domain" basis, which means that each element of the discretization is handled separately from the others, and a conciliation operation known as direct-stiffness summation is performed regularly to ensure continuity. 
This feature provides a great flexibility in preparing the data for post-processing, as such data analysis can often be performed locally without the need for additional communication.

\begin{figure}[t]
	\centering
\resizebox{0.48\textwidth}{!}{%
	\input{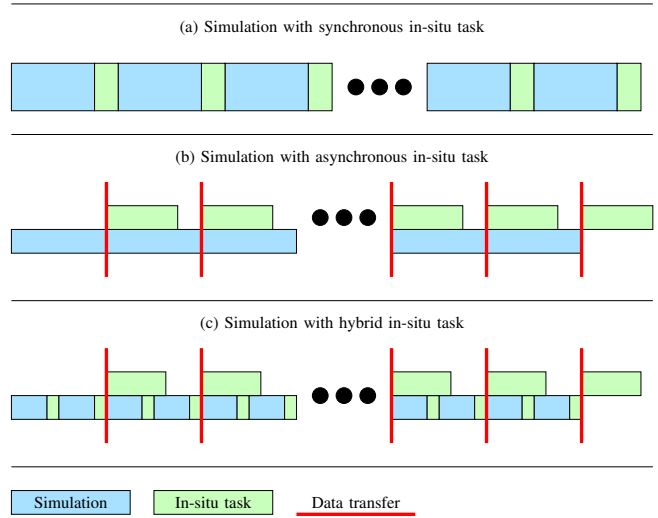}
}%
\vspace{-0.5em}
\caption{Illustration of a simulation with synchronous, asynchronous and hybrid in-situ tasks.}
\vspace{-1.5em}
\label{concept}
\end{figure}

CFD applications can fully utilize the computational power of Exascale High Performance Computing (HPC) systems with optimized data structure and parallelization. However, not only is the computational cost for CFD high but also is the amount of data generated, which grows with the size of the problem and need to be stored via the input/output (IO) subsystem for subsequent analysis (visual or numerical) as well as for checkpoint/restart operations. While the computational performance of HPC systems is rapidly increasing, the corresponding IO performance grows more slowly, and storage capacity is also limited. So the large IO operations, typical of the standard CFD workflows, may result in significant overheads. 

An alternative to reduce or prevent this overhead, is to perform analysis while the
simulation is running and the data resides in the HPC system's memory.
This type of approach is known as an {\it in-situ} approach~\cite{childs2020terminology}. In-situ techniques can reduce IO throughput and storage requirements, while improving overall simulation and data analysis performance. However, because computational resources must be shared between simulation and in-situ tasks, in-situ approaches may introduce new overheads. As a result, before deploying in-situ methods, the trade-off between reduced IO requirements and increased workloads must be carefully considered. 

Three main approaches can be identified based on how resources are shared and synchronized between simulation and in-situ tasks. In-situ tasks can be performed in a synchronous, asynchronous, or hybrid manner, the latter combining synchronous and asynchronous approaches as shown in Fig.~\ref{concept}.

In the {\it synchronous} in-situ approach, the simulation stops at well-defined intervals, allowing the in-situ task to run, and resumes after the end of the in-situ task (cf. Fig~\ref{concept}(a)). The in-situ task typically uses the same resources as the simulation, and by using appropriate data layouts, data copying can often be avoided, ideally, the data will still reside in the cache, making the in-situ processing very efficient. However, if the in-situ task requires a different data layout than the simulation, the caching may be destroyed, and the performance subsequently reduced. Due to overheads in the in-situ task, using all resources may be suboptimal, depending on the scalability of the in-situ task. It can also be difficult to decouple the in-situ task from the simulation because it relies on simulation functions and data structures, making it tightly coupled with the simulation and thus requiring synchronous execution. 

As shown in Fig.~\ref{concept}(b), the {\it asynchronous} in-situ approach uses separate computational resources for the simulation and in-situ task, allowing the simulation to transfer the data to the in-situ resources and then continue with the simulation while the in-situ task runs concurrently on the other resources. This allows the simulation and in-situ task to be decoupled, with a suitable amount of resources assigned to the in-situ task. However, it usually necessitates additional data copying, and determining how to divide available resources between simulation and in-situ tasks can be difficult. The suboptimal choice could result in load imbalances and consequently unnecessary waiting times. 

The {\it hybrid} in-situ approach (cf. Fig.~\ref{concept}(c)) combines the two, with the in-situ task divided into two parts. The first part is typically executed synchronously, followed by an asynchronous second part. In this approach, the synchronous part is executed on the same resources as the simulation (which pauses for the duration of the in-situ task), and then the required data are sent to the separate resources for the asynchronous part, after which the simulation can resume. While it is more difficult to design, it can overcome the disadvantages of the previous approaches by allowing sub-tasks that benefit from synchronous or asynchronous execution in their preferred model. 

As preciously discussed, these three in-situ approaches have different advantages and disadvantages, making it difficult to choose the best approach based on the characteristics of the simulation and in-situ tasks. 


In this paper, we investigate the impact of these three approaches on three common, yet very different in-situ tasks in CFD (compression of checkpoint/restart files, visualization, and uncertainty quantification) using a real CFD use case at scale (turbulent flow in a bent pipe \cite{hufnagel}). The following are the paper's specific contributions:
\begin{enumerate}
	\item it proposes both structures and examples for combining synchronous, asynchronous, and hybrid in-situ tasks and simulation;
	\item it presents three novel real-world case studies of in-situ tasks to large-scale simulation on CPU systems, including a new physics-based lossy data compression method; 
	\item it analyzes critically  which in-situ approach adds the least overhead to the simulation and achieves the best overall performance, generating experimental evidence for future model-based approaches. 
	
\end{enumerate}

The rest of the paper is organized as follows: Section~\ref{sec:back_relate} contains a summary of related works on in-situ techniques and case studies; Section~\ref{sec:method} introduces the paper's selected in-situ workflows and use cases; Section~\ref{sec:exp} contains information about the experimental setups; Section~\ref{sec:result} presents results, and analyses; Section~\ref{sec:conclude} summarizes and discusses this paper. 

\section{Related Work}
\label{sec:back_relate}

In-situ processing is gaining popularity, particularly in visualization and analysis, and several in-situ systems have been developed. VisIt with Libsim~\cite{childs2012visit,kuhlen2011parallel} and ParaView with Catalyst~\cite{ayachit2015paraview} are two in-situ systems for synchronous data visualization. SENSEI~\cite{ayachit2016sensei} is a generic in-situ interface that provides adaptors for connecting simulation to other in-situ systems like VisIt with Libsim and ParaView with Catalyst. It supports synchronous as well as asynchronous in-situ data analysis. Because these systems rely on the Visualization Toolkit (VTK) data format~\cite{schroeder1998visualization}, they can barely be used for tasks other than visualization.  

Originally designed as a higher-level IO abstraction, the Adaptable IO System (ADIOS)~\cite{liu2014hello, godoy2020adios} can also be used for in-situ processing. It is not dependent on VTK and supports arbitrary data formats, making it an excellent candidate for a generic in-situ framework. It also supports both synchronous and asynchronous in-situ tasks. As a result, we selected ADIOS as the framework for the work described in this paper. 

Several papers discuss the use of in-situ processing (mostly for visualization purposes) in CFD applications. \emph{Maulik et al.}~\cite{maulik2021pythonfoam}  evaluated the performance and scalability of three cases of OpenFOAM simulation with PythonFOAM performing Python-based synchronous data analysis. \emph{Ayachit et al.}~\cite{ayachit2016performance} visualized the simulation results from the PHASTA science application synchronously with the SENSEI generic interface in C/C++. Another group of researchers compared various in-situ approaches: \emph{Kress et al.}~\cite{kress2019comparing} compared the performance of synchronous and asynchronous data visualization in conjunction with a hydrodynamics proxy application that solved the compressible Euler equations. \emph{Oldfield et al.}~\cite{oldfield2014evaluation} applied in-situ data visualization to a large-scale shock physics code. They compared the performance and scalability of in-situ data visualization, both synchronous and asynchronous, with the traditional post-processing method. \emph{Bennett et al.}~\cite{bennett2012combining} proposed a hybrid in-situ approach for large-scale data analysis in a massively parallel turbulent combustion code (S3D) and compared the performance of their hybrid approach with a synchronous approach.

In contrast to these studies, we examine the suitability of all three in-situ approaches, synchronous, asynchronous, and hybrid, on large HPC systems using real large-scale CFD use cases and critically discuss their suitability based on the characteristics of the in-situ tasks. 

\begin{figure}[t]
	\centering
\resizebox{0.48\textwidth}{!}{%
	\begin{tikzpicture}
	\begin{pgfonlayer}{nodelayer}
		\node [style=none] (0) at (-5.25, 6) {};
		\node [style=none] (1) at (-7.75, 6) {};
		\node [style=none] (3) at (-4.75, 5.5) {};
		\node [style=none] (4) at (-4.75, 3.25) {};
		\node [style=none] (5) at (-5.25, 2.75) {};
		\node [style=none] (7) at (-7.75, 2.75) {};
		\node [style=none] (8) at (-8.25, 3.25) {};
		\node [style=none] (9) at (-8.25, 5.5) {};
		\node [style=none] (10) at (-1.25, 6) {};
		\node [style=none] (11) at (-3.75, 6) {};
		\node [style=none] (12) at (-0.75, 5.5) {};
		\node [style=none] (13) at (-0.75, 3.25) {};
		\node [style=none] (14) at (-1.25, 2.75) {};
		\node [style=none] (15) at (-3.75, 2.75) {};
		\node [style=none] (16) at (-4.25, 3.25) {};
		\node [style=none] (17) at (-4.25, 5.5) {};
		\node [style=none] (18) at (2.75, 6) {};
		\node [style=none] (19) at (0.25, 6) {};
		\node [style=none] (20) at (3.25, 5.5) {};
		\node [style=none] (21) at (3.25, 3.25) {};
		\node [style=none] (22) at (2.75, 2.75) {};
		\node [style=none] (23) at (0.25, 2.75) {};
		\node [style=none] (24) at (-0.25, 3.25) {};
		\node [style=none] (25) at (-0.25, 5.5) {};
		\node [style=none] (26) at (-6.5, 5.5) {Nek5000};
		\node [style=blueLargeRec] (38) at (-6.5, 4.75) {$in\_situ\_init$};
		\node [style=blueLargeRec] (39) at (-6.5, 4) {$in\_situ\_check$};
		\node [style=blueLargeRec] (40) at (-6.5, 3.25) {$in\_situ\_end$};
		\node [style=none] (41) at (-2.5, 5.5) {Nek-proc adaptor};
		\node [style=none] (42) at (1.5, 5.5) {Data proccessor};
		\node [style=redLargeRec] (43) at (-2.5, 4.75) {$nek\_proc\_init$};
		\node [style=redLargeRec] (44) at (-2.5, 4) {$nek\_proc\_check$};
		\node [style=redLargeRec] (45) at (-2.5, 3.25) {$nek\_proc\_end$};
		\node [style=none] (46) at (-5.25, 4.75) {};
		\node [style=none] (47) at (-4, 4.75) {};
		\node [style=none] (48) at (-5.25, 4) {};
		\node [style=none] (49) at (-4, 4) {};
		\node [style=none] (50) at (-5.25, 3.25) {};
		\node [style=none] (51) at (-4, 3.25) {};
		\node [style=blueLargeRec] (61) at (-6.5, 6.5) {Simulation subroutines};
		\node [style=greenLargeRec] (62) at (1.5, 6.5) {Data processor functions};
		\node [style=greenLargeRec] (63) at (1.5, 4.75) {$processor\_init$};
		\node [style=greenLargeRec] (64) at (1.5, 3.25) {$processor\_end$};
		\node [style=greenLargeRec] (65) at (1.5, 4) {$processor\_check$};
		\node [style=none] (66) at (-1.25, 4.75) {};
		\node [style=none] (67) at (0, 4.75) {};
		\node [style=none] (68) at (-1.25, 4) {};
		\node [style=none] (69) at (0, 4) {};
		\node [style=none] (70) at (-1.25, 3.25) {};
		\node [style=none] (71) at (0, 3.25) {};
		\node [style=redLargeRec] (72) at (-2.5, 6.5) {Adaptor functions};
	\end{pgfonlayer}
	\begin{pgfonlayer}{edgelayer}
		\draw [style=myEdge] (8.center)
			 to (9.center)
			 to [bend left=45, looseness=1.75] (1.center)
			 to [in=180, out=0] (0.center)
			 to [bend left=45, looseness=1.75] (3.center)
			 to (4.center)
			 to [bend left=45, looseness=1.75] (5.center)
			 to (7.center)
			 to [bend right=315, looseness=1.50] cycle;
		\draw [style=myEdge] (16.center)
			 to (17.center)
			 to [bend left=45, looseness=1.75] (11.center)
			 to (10.center)
			 to [bend left=45, looseness=1.75] (12.center)
			 to (13.center)
			 to [bend left=45, looseness=1.75] (14.center)
			 to (15.center)
			 to [bend right=315, looseness=1.50] cycle;
		\draw [style=myEdge] (24.center) to (25.center);
		\draw [style=myEdge, bend left=45, looseness=1.75] (25.center) to (19.center);
		\draw [style=myEdge] (19.center) to (18.center);
		\draw [style=myEdge, bend left=45, looseness=1.75] (18.center) to (20.center);
		\draw [style=myEdge] (20.center) to (21.center);
		\draw [style=myEdge, bend left=45, looseness=1.75] (21.center) to (22.center);
		\draw [style=myEdge] (22.center) to (23.center);
		\draw [style=myEdge, bend right=315, looseness=1.50] (23.center) to (24.center);
		\draw [style=myEdgeArrows] (46.center) to (47.center);
		\draw [style=myEdgeArrows] (48.center) to (49.center);
		\draw [style=myEdgeArrows] (50.center) to (51.center);
		\draw [style=myEdgeArrows] (66.center) to (67.center);
		\draw [style=myEdgeArrows] (68.center) to (69.center);
		\draw [style=myEdgeArrows] (70.center) to (71.center);
	\end{pgfonlayer}
\end{tikzpicture}
}%
\vspace{-0.5em}
\caption{Illustration of the workflow of a Nek5000 simulation with a synchronous in-situ task.}
\label{sync}
\vspace{-1.5em}
\end{figure}
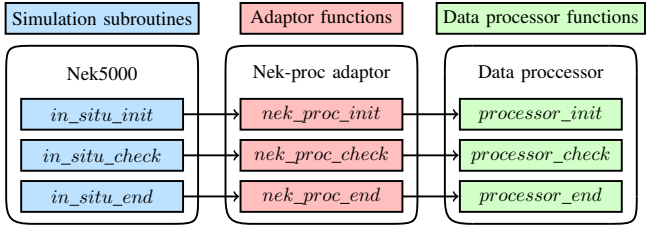

\section{Methodology}
\label{sec:method}

We use the incompressible spectral-element Navier-Stokes solver, Nek5000, and three common in-situ tasks to investigate the impact of in-situ tasks on CFD simulations: lossy and lossless compression, data visualization with image generation, and data analysis for uncertainty quantification. In this section, we will present our synchronous, asynchronous and hybrid in-situ workflow and introduce these three uses cases.

\subsection{In-Situ Workflow}

Both asynchronous and hybrid in-situ approaches require data transfer from the simulation to the in-situ tasks, and rather than using custom communication approaches, we use the ADIOS2 \cite{godoy2020adios} framework, which provides APIs in Fortran, C/C++, and Python as well as several in-situ functions, including the ``\emph{insituMPI}" engine for MPI programs, thus providing a stable framework for similar approaches in other contexts. 
As previously stated, 
ADIOS2 does not rely on the VTK data format, allowing for easier integration and avoiding the need for additional data copying to the VTK format. 

\begin{figure}[t]
	\centering
\resizebox{0.48\textwidth}{!}{%
	\begin{tikzpicture}
	\begin{pgfonlayer}{nodelayer}
		\node [style=none] (0) at (-5.25, 6) {};
		\node [style=none] (1) at (-7.75, 6) {};
		\node [style=none] (3) at (-4.75, 5.5) {};
		\node [style=none] (4) at (-4.75, 3.25) {};
		\node [style=none] (5) at (-5.25, 2.75) {};
		\node [style=none] (7) at (-7.75, 2.75) {};
		\node [style=none] (8) at (-8.25, 3.25) {};
		\node [style=none] (9) at (-8.25, 5.5) {};
		\node [style=none] (10) at (-1.25, 6) {};
		\node [style=none] (11) at (-3.75, 6) {};
		\node [style=none] (12) at (-0.75, 5.5) {};
		\node [style=none] (13) at (-0.75, 3.25) {};
		\node [style=none] (14) at (-1.25, 2.75) {};
		\node [style=none] (15) at (-3.75, 2.75) {};
		\node [style=none] (16) at (-4.25, 3.25) {};
		\node [style=none] (17) at (-4.25, 5.5) {};
		\node [style=none] (18) at (-5.25, 2) {};
		\node [style=none] (19) at (-7.75, 2) {};
		\node [style=none] (20) at (-4.75, 1.5) {};
		\node [style=none] (21) at (-4.75, -0.75) {};
		\node [style=none] (22) at (-5.25, -1.25) {};
		\node [style=none] (23) at (-7.75, -1.25) {};
		\node [style=none] (24) at (-8.25, -0.75) {};
		\node [style=none] (25) at (-8.25, 1.5) {};
		\node [style=none] (26) at (-6.5, 5.5) {Nek5000};
		\node [style=blueLargeRec] (38) at (-6.5, 4.75) {$in\_situ\_init$};
		\node [style=blueLargeRec] (39) at (-6.5, 4) {$in\_situ\_check$};
		\node [style=blueLargeRec] (40) at (-6.5, 3.25) {$in\_situ\_end$};
		\node [style=none] (41) at (-2.5, 5.5) {Nek-writer adaptor};
		\node [style=none] (42) at (-6.5, 1.5) {Data proccessor};
		\node [style=redLargeRec] (43) at (-2.5, 4.75) {$nek\_wrtr\_init$};
		\node [style=redLargeRec] (44) at (-2.5, 4) {$nek\_wrtr\_write$};
		\node [style=redLargeRec] (45) at (-2.5, 3.25) {$nek\_wrtr\_end$};
		\node [style=none] (46) at (-5.25, 4.75) {};
		\node [style=none] (47) at (-4, 4.75) {};
		\node [style=none] (48) at (-5.25, 4) {};
		\node [style=none] (49) at (-4, 4) {};
		\node [style=none] (50) at (-5.25, 3.25) {};
		\node [style=none] (51) at (-4, 3.25) {};
		\node [style=blueLargeRec] (61) at (-6.5, 7.25) {Simulation subroutines};
		\node [style=greenLargeRec] (62) at (1.5, 7.25) {Data processor functions};
		\node [style=greenLargeRec] (63) at (-6.5, 0.75) {$processor\_init$};
		\node [style=greenLargeRec] (64) at (-6.5, -0.75) {$processor\_end$};
		\node [style=greenLargeRec] (65) at (-6.5, 0) {$processor\_check$};
		\node [style=none] (66) at (-1.25, 4.75) {};
		\node [style=none] (67) at (0, 4.75) {};
		\node [style=none] (68) at (-1.25, 4) {};
		\node [style=none] (69) at (0, 4) {};
		\node [style=none] (70) at (-1.25, 3.25) {};
		\node [style=none] (71) at (0, 3.25) {};
		\node [style=redLargeRec] (72) at (-2.5, 7.25) {Adaptor functions};
		\node [style=none] (73) at (2.75, 6) {};
		\node [style=none] (74) at (0.25, 6) {};
		\node [style=none] (75) at (3.25, 5.5) {};
		\node [style=none] (76) at (3.25, 3.25) {};
		\node [style=none] (77) at (2.75, 2.75) {};
		\node [style=none] (78) at (0.25, 2.75) {};
		\node [style=none] (79) at (-0.25, 3.25) {};
		\node [style=none] (80) at (-0.25, 5.5) {};
		\node [style=none] (81) at (1.5, 5.5) {ADIOS writer};
		\node [style=yellowLargRec] (82) at (1.5, 4.75) {$writer\_init$};
		\node [style=yellowLargRec] (83) at (1.5, 3.25) {$writer\_end$};
		\node [style=yellowLargRec] (84) at (1.5, 4) {$writer\_check$};
		\node [style=none] (85) at (-1.25, 2) {};
		\node [style=none] (86) at (-3.75, 2) {};
		\node [style=none] (87) at (-0.75, 1.5) {};
		\node [style=none] (88) at (-0.75, -0.75) {};
		\node [style=none] (89) at (-1.25, -1.25) {};
		\node [style=none] (90) at (-3.75, -1.25) {};
		\node [style=none] (91) at (-4.25, -0.75) {};
		\node [style=none] (92) at (-4.25, 1.5) {};
		\node [style=none] (93) at (-2.5, 1.5) {Reader-proc adaptor};
		\node [style=redLargeRec] (94) at (-2.5, 0.75) {$rdr\_proc\_init$};
		\node [style=redLargeRec] (95) at (-2.5, 0) {$rdr\_proc\_check$};
		\node [style=redLargeRec] (96) at (-2.5, -0.75) {$rdr\_proc\_end$};
		\node [style=none] (97) at (-5, 0.75) {};
		\node [style=none] (98) at (-3.75, 0.75) {};
		\node [style=none] (99) at (-5, 0) {};
		\node [style=none] (100) at (-3.75, 0) {};
		\node [style=none] (101) at (-5, -0.75) {};
		\node [style=none] (102) at (-3.75, -0.75) {};
		\node [style=none] (103) at (-1, 0.75) {};
		\node [style=none] (104) at (0.25, 0.75) {};
		\node [style=none] (105) at (-1, 0) {};
		\node [style=none] (106) at (0.25, 0) {};
		\node [style=none] (107) at (-1, -0.75) {};
		\node [style=none] (108) at (0.25, -0.75) {};
		\node [style=none] (109) at (2.75, 2) {};
		\node [style=none] (110) at (0.25, 2) {};
		\node [style=none] (111) at (3.25, 1.5) {};
		\node [style=none] (112) at (3.25, -0.75) {};
		\node [style=none] (113) at (2.75, -1.25) {};
		\node [style=none] (114) at (0.25, -1.25) {};
		\node [style=none] (115) at (-0.25, -0.75) {};
		\node [style=none] (116) at (-0.25, 1.5) {};
		\node [style=none] (117) at (1.5, 1.5) {ADIOS reader};
		\node [style=yellowLargRec] (118) at (1.5, 0.75) {$reader\_init$};
		\node [style=yellowLargRec] (119) at (1.5, -0.75) {$ender\_end$};
		\node [style=yellowLargRec] (120) at (1.5, 0) {$reader\_check$};
		\node [style=none] (122) at (1.5, 2.75) {};
		\node [style=none] (125) at (1.5, 2) {};
		\node [style=none] (128) at (3, 4.75) {};
		\node [style=none] (130) at (3, 3.25) {};
		\node [style=yellowLargRec] (132) at (-6.5, 6.5) {ADIOS functions};
		\node [style=none] (133) at (-3.75, 6.5) {};
		\node [style=none] (134) at (-3, 6.5) {};
		\node [style=none] (135) at (-1.5, 6.5) {MPI data transfer};
	\end{pgfonlayer}
	\begin{pgfonlayer}{edgelayer}
		\draw [style=myEdge] (8.center)
			 to (9.center)
			 to [bend left=45, looseness=1.75] (1.center)
			 to [in=180, out=0] (0.center)
			 to [bend left=45, looseness=1.75] (3.center)
			 to (4.center)
			 to [bend left=45, looseness=1.75] (5.center)
			 to (7.center)
			 to [bend right=315, looseness=1.50] cycle;
		\draw [style=myEdge] (16.center)
			 to (17.center)
			 to [bend left=45, looseness=1.75] (11.center)
			 to (10.center)
			 to [bend left=45, looseness=1.75] (12.center)
			 to (13.center)
			 to [bend left=45, looseness=1.75] (14.center)
			 to (15.center)
			 to [bend right=315, looseness=1.50] cycle;
		\draw [style=myEdge] (24.center) to (25.center);
		\draw [style=myEdge, bend left=45, looseness=1.75] (25.center) to (19.center);
		\draw [style=myEdge] (19.center) to (18.center);
		\draw [style=myEdge, bend left=45, looseness=1.75] (18.center) to (20.center);
		\draw [style=myEdge] (20.center) to (21.center);
		\draw [style=myEdge, bend left=45, looseness=1.75] (21.center) to (22.center);
		\draw [style=myEdge] (22.center) to (23.center);
		\draw [style=myEdge, bend right=315, looseness=1.50] (23.center) to (24.center);
		\draw [style=myEdgeArrows] (46.center) to (47.center);
		\draw [style=myEdgeArrows] (48.center) to (49.center);
		\draw [style=myEdgeArrows] (50.center) to (51.center);
		\draw [style=myEdgeArrows] (66.center) to (67.center);
		\draw [style=myEdgeArrows] (68.center) to (69.center);
		\draw [style=myEdgeArrows] (70.center) to (71.center);
		\draw [style=myEdge] (79.center) to (80.center);
		\draw [style=myEdge, bend left=45, looseness=1.75] (80.center) to (74.center);
		\draw [style=myEdge] (74.center) to (73.center);
		\draw [style=myEdge, bend left=45, looseness=1.75] (73.center) to (75.center);
		\draw [style=myEdge] (75.center) to (76.center);
		\draw [style=myEdge, bend left=45, looseness=1.75] (76.center) to (77.center);
		\draw [style=myEdge] (77.center) to (78.center);
		\draw [style=myEdge, bend right=315, looseness=1.50] (78.center) to (79.center);
		\draw [style=myEdge] (91.center)
			 to (92.center)
			 to [bend left=45, looseness=1.75] (86.center)
			 to (85.center)
			 to [bend left=45, looseness=1.75] (87.center)
			 to (88.center)
			 to [bend left=45, looseness=1.75] (89.center)
			 to (90.center)
			 to [bend right=315, looseness=1.50] cycle;
		\draw [style=myEdge2] (97.center) to (98.center);
		\draw [style=myEdge2] (99.center) to (100.center);
		\draw [style=myEdge2] (101.center) to (102.center);
		\draw [style=myEdge2] (103.center) to (104.center);
		\draw [style=myEdge2] (105.center) to (106.center);
		\draw [style=myEdge2] (107.center) to (108.center);
		\draw [style=myEdge] (115.center) to (116.center);
		\draw [style=myEdge, bend left=45, looseness=1.75] (116.center) to (110.center);
		\draw [style=myEdge] (110.center) to (109.center);
		\draw [style=myEdge, bend left=45, looseness=1.75] (109.center) to (111.center);
		\draw [style=myEdge] (111.center) to (112.center);
		\draw [style=myEdge, bend left=45, looseness=1.75] (112.center) to (113.center);
		\draw [style=myEdge] (113.center) to (114.center);
		\draw [style=myEdge, bend right=315, looseness=1.50] (114.center) to (115.center);
		\draw [style=dataflow] (122.center) to (125.center);
		\draw [style=dataflow] (133.center) to (134.center);
	\end{pgfonlayer}
\end{tikzpicture}
}%
\vspace{-0.5em}
\caption{Illustration of the workflow of a Nek5000 simulation with an asynchronous in-situ task.}
\vspace{-1.5em}
\label{async}
\end{figure}
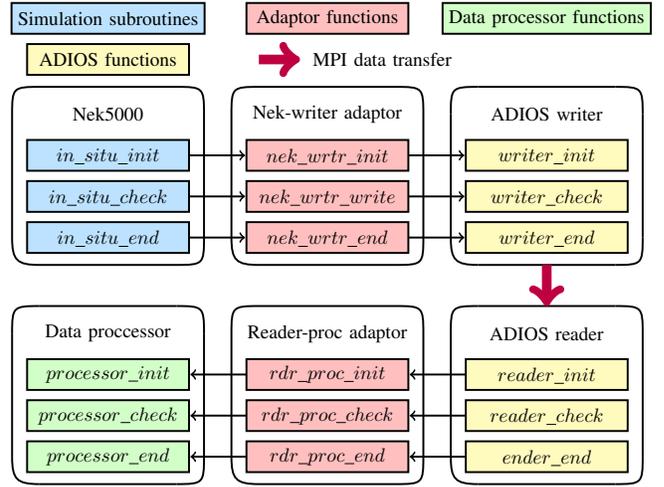

\begin{figure*}[t]
	\centering
\resizebox{0.8\textwidth}{!}{%
	\input{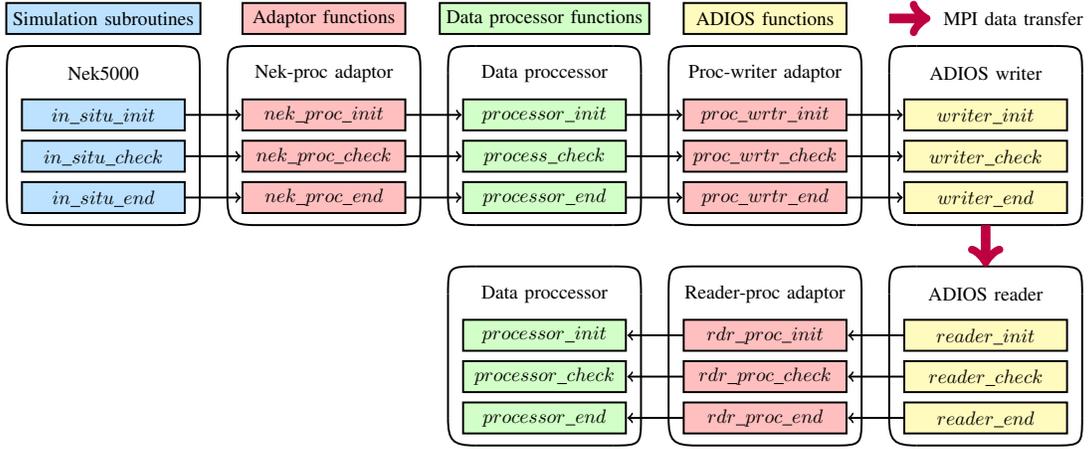}
}%
\vspace{-0.5em}
\caption{Illustration of the workflow of a Nek5000 simulation with a hybrid in-situ task.}
\vspace{-1.5em}
\label{hybrid}
\end{figure*}

The first technical issue for in-situ approaches is compiling and linking the simulation codes with the in-situ tasks; this is frequently difficult because the simulation and in-situ tasks may be programmed in different languages. To address this issue, we use adaptor functions in our workflow design (Fig.~\ref{sync}-\ref{hybrid}), which are wrappers written in the programming languages used to code the simulation and/or in-situ task. 

To enable in-situ processing, three groups of functions must be implemented in the simulation: \textit{initialization}, \textit{check} and \textit{finalization}. These functions connect the in-situ tasks to the simulation solver, Nek5000. The \textit{initialization} is implemented with the $*\_init$ functions. During the initialization phase, the simulation solver and in-situ task call them. These functions set necessary parameters, such as the size of the data to be processed or transferred during the in-situ step, based on information from the simulation solver and the in-situ task. The \textit{finalization} is accomplished with the $*\_end$ functions. These finalize the in-situ setup safely, free up all used memory, and print out profiling information on the in-situ tasks. The \textit{Check} is a group of $*\_check$ functions that are called in each step when the in-situ task is executed. 

In the synchronous approach, the simulation results are passed to the in-situ task via the adaptor functions, as shown in Fig.~\ref{sync}. The simulation is stopped during in-situ execution, thus data consistency is guaranteed automatically. If the in-situ task and simulation use the same data structures, no additional data copying is required. If, on the other hand, the in-situ task requires a different data structure, the adaptor functions may need to include a deep copy. 
Because the simulation solver and the in-situ task share computing resources, the core used to perform the in-situ task already has the simulation results. As a result, we do not need to use the ADIOS2 library to transfer data between cores. 

In the asynchronous approach, the simulation results are sent to the in-situ task via the writer and reader pair based on the ``\emph{insituMPI}" engine from ADIOS2 (Fig.~\ref{async}). The simulation solver and the in-situ task need to be launched concurrently in a multiple-program multiple-data (MPMD) mode. The simulation and in-situ task workloads are distributed across separate computational resources. Given the total number of resources (e.g. cores) available, $N$, they can be assigned in various chunks to the simulation $p_{sim}$ and in-situ task $p_{insitu}$ such that $p_{sim}+p_{insitu}=N$. 
The simulation sends the required data to the ADIOS2 writer via adaptor functions. To ensure data consistency, the simulation waits for the end of the MPI communication. The in-situ task is executed concurrently with the simulation after receiving the data from the ADIOS2 reader. If the simulation solver and the data processor have different structures, the adaptor functions that connect the data processor and the reader perform the necessary adaptations. 

The hybrid in-situ approach depicted in Fig.~\ref{hybrid} is divided into synchronous and asynchronous components. The adaptor functions, like the synchronous approach, pass the simulation results to the first synchronous part of the in-situ task. Following the synchronous portion, intermediate data is sent to the second portion of the in-situ task via ADIOS2, as in the asynchronous approach. The simulation solver is directly compiled and linked with the synchronous part of the in-situ task in this approach, and it is launched in MPMD mode with the asynchronous part of the in-situ task.

\subsection{Use Cases}
We consider three use cases with various characteristics from large CFD simulations to assess the efficiency of our in-situ approaches: 

\paragraph{\textbf{Lossy and lossless compression}} 
CFD simulations are frequently long-running, producing potentially large amounts of output data for post-processing or check-pointing/restart mechanisms. Compressing the data before storing it is one way to reduce the storage requirements of a simulation. According to \emph{Li et al.}~\cite{Li}, many types of compression can be applied to data sets, but in this study we only distinguish between \textbf{lossless  compression}, where no information from the original data is discarded, and \textbf{lossy  compression} where there is no demand for the reconstructed data set to exactly match the original one, introducing errors but allowing for higher compression ratios. Ideally, lossless compression may be preferred in all cases, as scientists prefer to have undisturbed data for any necessary analysis. However, turbulence is characterized by seemingly random motions, which add a level of complexity for lossless encoders that typically rely on finding patterns in the data, reducing their ability to perform significant compression. As a result, lossy compression is widely regarded as a viable alternative. Turbulence is a chaotic multi-scale phenomenon in which there are motions at various frequencies, but only a few of them ultimately possess the majority of the energy in the flow. It is possible to keep only the data associated with the most energetic motions in the flow while discarding the rest using a method proposed by \emph{Otero et al.}~\cite{otero2018lossy}. This allows for the data to be reconstructed with reasonable accuracy. Another advantage of this physics-based method is that the user can specify allowed error in the reconstructed data set in advance, and compression will take place element wise accordingly.

This physics-based method is inspired by the JPEG compression standard~\cite{JPEG}, with the exception that it employs the Discrete Legendre Transform (DLT) rather than the Discrete Cosine Transform (DCT). This specific transformation is chosen in order to benefit from the Gauss-Lobatto-Legendre (GLL) points that are used in the spectral element discretization \cite{deville_fischer_mund_2002}. The DLT is used to transform the original data into spectral space, and low energy spectral coefficients are systematically discarded in an inherently lossy step known as \textbf{truncation}. While the truncation is taking place, we use the orthogonality and other properties of the Legendre basis to evaluate the error incurred on the original data set without transforming back into the physical domain, lowering the computational cost of the method. Lossy compression is completed by using Lossless Huffman encoding or another suitable method and writing the data to disk. 
For the latter task, we use the functions for lossless compression and IO from the ADIOS2 library as part of the in-situ data processor even in the synchronous approach because the data compression is a special in-situ task. The ADIOS2 library is not used in the synchronous approach in the later cases.

Ideally, the data compression should take relatively short time since it is a fully local operation, and could compress the data to a certain degree while keeping the sufficient accuracy. Because of this, we have chosen this case as one example of in-situ tasks with low computational cost. And due to the reusage of the simulation functions of the lossy compression, this use case is also one example of in-situ tasks, which is partly deep coupled with the simulation. 


\paragraph{\textbf{Visualization with ParaView/Catalyst}} In-situ visualization can eliminate the need for intermediate data storage for the visualization (often postmortem), improving the overall simulation and  visualization efficiency. However, because of the required collective communications, the visualization task frequently scales much less than the simulation. As a result, synchronous in-situ approaches can be problematic with the MPI collective communication, as shown by \emph{Atzori et al.}~\cite{atzori2022situ}. Using the  asynchronous in-situ approach, it is thus advantageous to assign a smaller set of resources to the in-situ task than to the simulation. We use ParaView/Catalyst as an image generator. The general image generation workflow can be expressed as follows: the VTK grid for ParaView-based visualization is generated during the initialization phase, and a customized ParaView Pipeline Python script is read. This Python script defines how the ParaView/Catalyst coprocessor renders the output image using information such as camera position, image size, and slice position.

According to the bottleneck diagnosed by the previous study, this image generation is one example of in-situ tasks with worse scalability compared to the simulation. 

\paragraph{\textbf{Data analysis -- uncertainty quantification (UQ)}} Uncertainty quantification is important for accessing the reliability of computed turbulence statistics, which are required to understand the relevant physics and formal analysis of turbulent flow simulations~\cite{Gscheidle2022framwork}. The UQ data analysis is the in-situ task in this case and is divided into two portions. The first portion is to update the  sample-estimated autocorrelation function at a series of time lags, known as training lags. The second portion is to use the sample-estimated values to model the autocorrelation function, and calculate the uncertainty in a sample mean. The first portion is executed more frequently than the second, but has a lower computational cost, i.e. the first portion takes only neglectable time compared to the second portion.  

Because of the different frequency and computational cost of individual portion of uncertainty quantification, it is an example of complex in-situ tasks with different portions, which are suitable to different of in-situ techniques. 


\section{Experimental Setup} 
\label{sec:exp}
We introduce the system setups, CFD case and evaluation metrics of our use case evaluation in this section.

\paragraph{\textbf{System setup}} We used two HPC systems, the Raven supercomputer at the Max Planck Computing and Data Facility (MPCDF)~\cite{Raven} and the Dardel supercomputer at the PDC Centre for High-Performance Computing (PDC-HPC) at Royal Institute of Technology (KTH)~\cite{Dardel}. 
One Raven node contains two Intel Xeon IceLake-SP 8360Y processors with 36 cores each and 256 GB RAM.
Each Dardel node contains two AMD EPYIC processors, each with 64 cores and 256 GB RAM. 

The MPMD configuration file defines how cores are allocated to simulation solvers and in-situ tasks for the asynchronous and hybrid in-situ approach. On each node, one set of cores is dedicated to simulation, while the rest are dedicated to the in-situ task. In this way, data transfer is only required on the node.  

\paragraph{\textbf{CFD Case}} We chose the turbulent flow inside a bent pipe, which is an internal flow, i.e., a flow bounded by walls, and exhibits many of the most critical turbulence characteristics. Additionally, the bent pipe exhibits low frequency dynamics in a phenomenon known as swirl switching \cite{hufnagel}, which makes it interesting to determine the effect of our in-situ techniques in more complex problems. We took precautions in all cases to ensure that turbulence has already developed when we apply the in-situ techniques. The CPU-based simulation uses a discretization of the physical domain with 459000 elements with accuracy of order seven, i.e 512 data points per element. A true Exascale simulation of turbulence would possess of the order of tens of millions of elements, however the work per processing element would remain similar to what used in the current simulations, thus we expect the behaviour to be transferable to larger cases.

\paragraph{\textbf{Evaluation metrics}} To evaluate the performance of the simulation with the in-situ tasks, we measure the execution time, perform profiling, and analyze scalability. As the performance metric, we measure the execution of 1000 simulation steps and evaluate the average execution time of one simulation step as the performance metric. For the synchronous approach, we perform the strong scalability test; for the asynchronous and hybrid approach, we first perform the configuration tests on fixed number of nodes, and repeat the these tests on different number of nodes. 

We analyze the compression ratios obtained as well as properly weighted root mean squared error (RMSE) of the reconstructed data set to evaluate the compression. We also investigate whether the reconstructed data fields could be used to create meaningful visualizations. To visually verify the results, we compare the images generated both synchronously and asynchronously with the images generated in post-processing from VTK files. We measure the size of the VTK files for the post-processing image generation and uncertainty quantification, which is not needed in the in-situ case, to demonstrate the memory savings from in-situ techniques. 

We repeat each experiment three times, and the arithmetic average of the obtained evaluation metrics is reported.

\section{Use case evaluations}
\label{sec:result}

In this section, we evaluate the CFD simulation with the three use cases described using synchronous, asynchronous, and hybrid in-situ executions.

\subsection{In-situ data compression}

We first study  synchronous and hybrid in-situ data compression to a turbulent fluid in a bent pipe simulation with Nek5000.

\paragraph{Implementation}
For \textit{synchronous data compression}, we reuse the Fortran functions from Nek5000 to execute the lossy physics-based truncation mentioned in the previous section, and use C/C++-based adaptor functions to pass the output to the C/C++ ADIOS2 writer, which subsequently performs lossless BZIP2 compression and IO operations. The \textsf{Nek-proc} adaptor functions in Fig.~\ref{sync} are written in Fortran in this case, and no communication is required because the cores already hold the data to perform lossless compression and to write out locally.

Because lossy compression is tightly coupled with the simulation, fully asynchronous compression is difficult to achieve. As a result, we tested  \textit{hybrid data compression}, in which we use the same functions, workflow and the \textsf{Nek-proc} adaptor functions (Fig.~\ref{hybrid}) as in the synchronous case, but asynchronously perform the lossless compression. We can use the C/C++ writer in the ADIOS2 writer-reader pair instead of the file writer with lossless compression from ADIOS2 with runtime configurations. The data is synchronously truncated in a lossy manner with the simulation and then passed through \textsf{Proc-writer} adaptor functions, which are a group of functions in Fortran and C/C++, to the ADIOS writer. Lossless data compression is then performed asynchronously, is entirely programmed in C/C++ and runs on a different set of cores from the original simulation. For this case, C/C++-based \textsf{Reader-proc} adaptor functions connect a reader in the ADIOS2 writer-reader pair to a separate file writer with lossless compression from ADIOS2. We should point out that the workload on each core for lossless compression is distributed evenly, which is not necessarily the case in the simulation.



\begin{figure}[t]
\centering
	\includegraphics[width=0.48\textwidth]{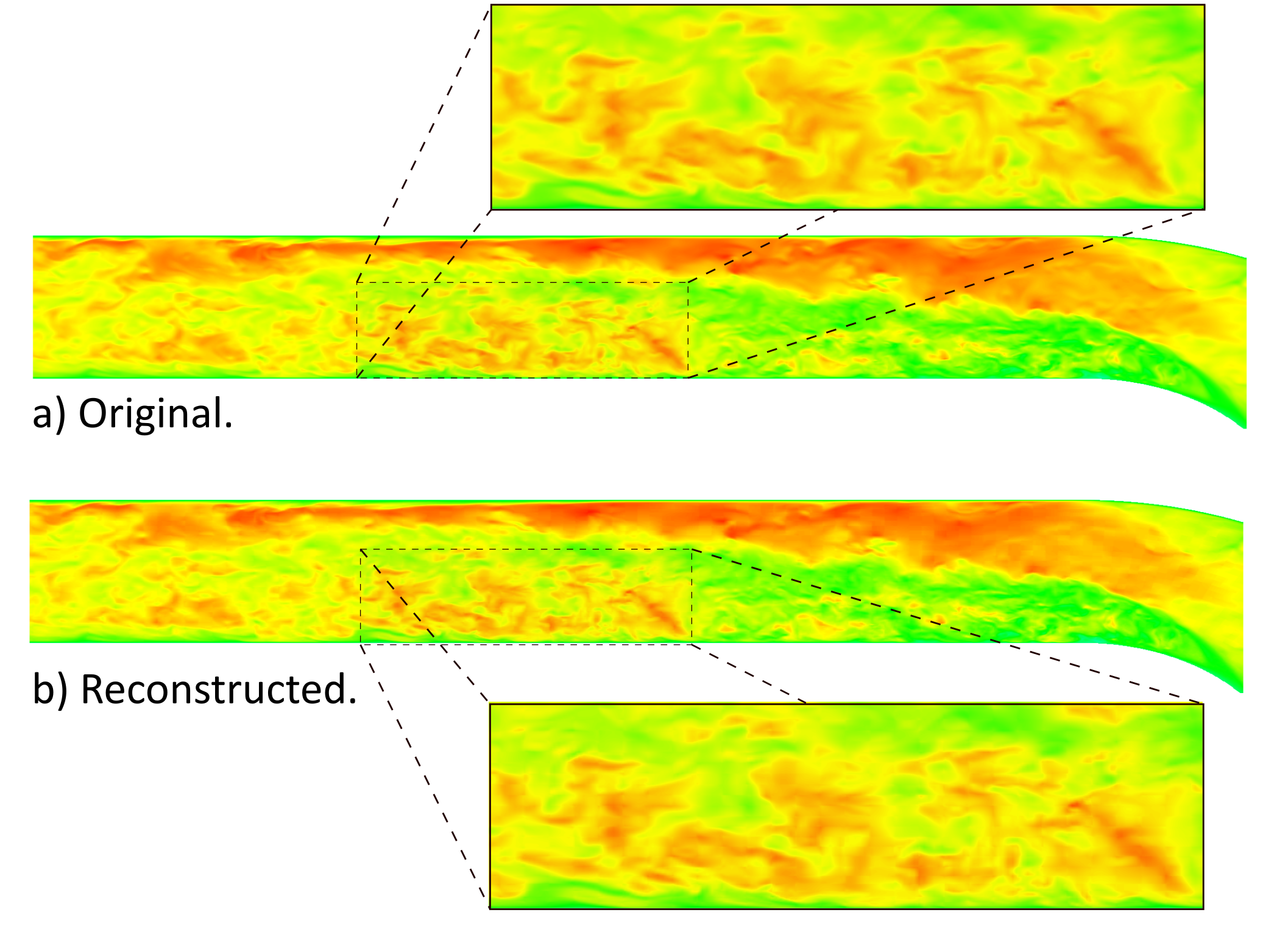}
	\vspace{-1em}
	\caption{Slice of the velocity magnitude downstream from the bent section. a) is the original data set, while b) is the reconstruction of a field compressed with a maximum allowed error of $10^{-2}$.}
	\vspace{-1em}
	\label{comp_viz}
\end{figure}

 \paragraph{Evaluation}
To show that compressed data sets, even at high compression ratios, are still relevant and meaningful for analysis, we present Fig.~\ref{comp_viz}, where we show a slice of one reconstructed velocity component for compression with input error $\epsilon = 10^{-2}$ , which correspond to a file with a compression ratio of 51, i.e $98\%$ the data has been discarded. In the figure we observe that all the features of the turbulent flow are preserved even at these rates.

\begin{figure*}[]
\centering
\includegraphics[clip,width=0.98\textwidth]{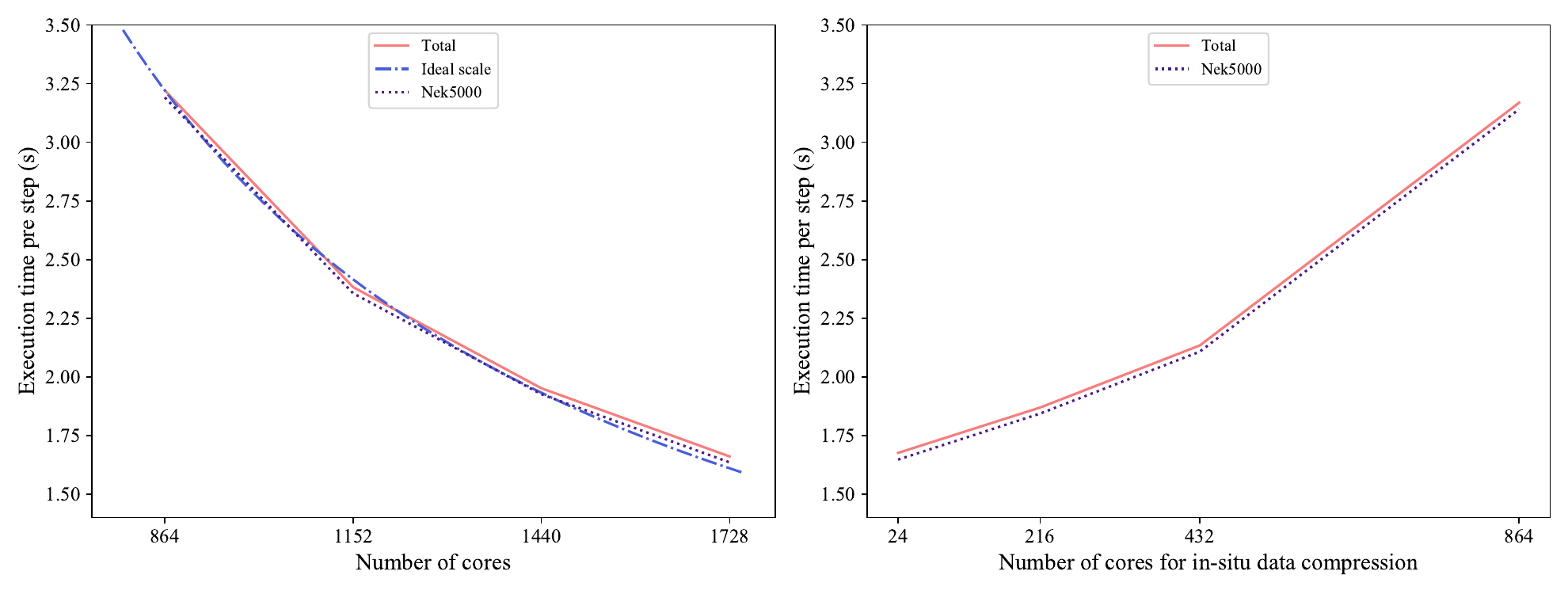}
\vspace{-1.5em}
\caption{Execution time of Nek5000 with synchronous in-situ compression with lossy compression maximum allowed error $\epsilon=10^{-2}$ on Raven supercomputer (left) and hybrid in-situ compression with lossy compression maximum allowed error $\epsilon=10^{-2}$ on 24 Raven nodes (right).}
\label{compre_res}
	\vspace{-1.em}
\end{figure*}

We observe that most compression artifacts happen at the element boundaries, mostly because for spectral element methods, continuity among elements is enforced weakly by using direct stiffness summation and the compression scheme we use truncates information locally at the element level without care for neighbouring data. However, this property allows for minimal communication and computation that ultimately produces the performance that will be subsequently shown in Fig. \ref{compre_res}.
We note that the compressed fields have been shown to produce correct statistics and modal decomposition even with the presence of such artifacts.~\cite{otero2018lossy,Marin}.

\begin{figure}[t]
\centering
\vspace{-1.em}
	\includegraphics[width=0.48\textwidth]{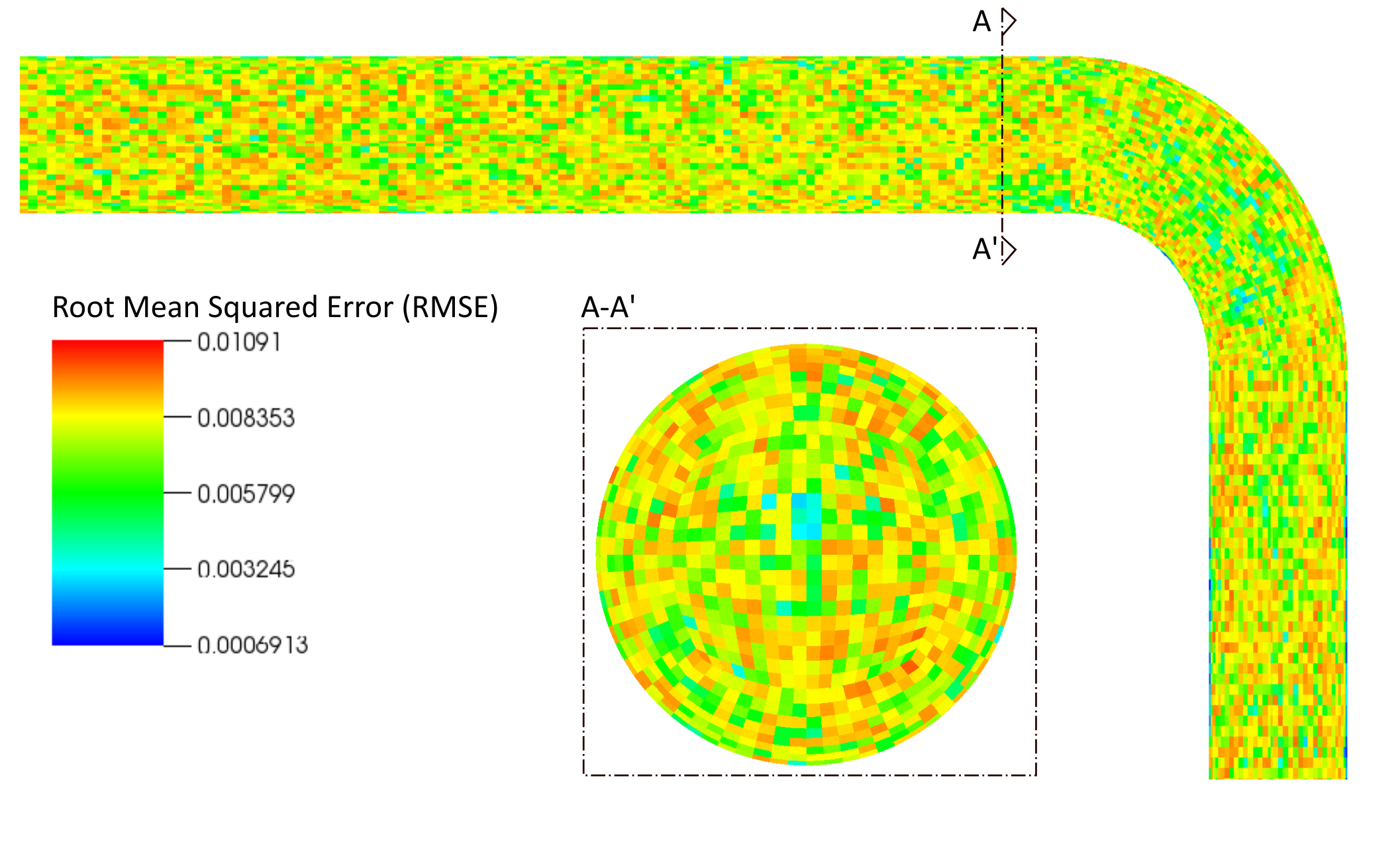}
	\vspace{-1.5em}
	\caption{RMSE of a slice of the 3D field for a maximum allowed error of $10^{-2}$. The error is shown per spectral element.}
	\label{comp_rmse}
	\vspace{-1em}
\end{figure}

In Fig.~\ref{comp_rmse}, we show the post-computed RMSE between the original and reconstructed fields in physical space for compression with the maximum allowed error of $\epsilon=10^{-2}$.
As expected, higher compression ratios can be attained by allowing higher errors and even if not explicitly calculated at run time with physical variables, the errors are always still within the appropriate bounds in the reconstructed data sets.

Having confirmed that no relevant artifacts are introduced due to the in-situ implementation of compression on both Dardel and Raven supercomputers, we analyze the performance of the implementation.

For this purpose we ran a strong scalability test for the simulation with synchronous in-situ data compression every 50 simulation steps, which is a high frequency to write checkpoint/restart file, using 12, 16, 20, and 24 nodes on Raven (i.e., 864, 1152, 1440 and 1728 cores). As shown in the left graph of Fig.~\ref{compre_res}, the execution time of Nek5000 with this configuration decreases as the number of cores increases, and it achieves excellent strong scalability. The execution of Nek5000 consumes the majority of the time, while the compression and data output consume a negligible part of the total time (1.5 \% of the total execution time). 

We further profiled the performance of lossless compression and data writing in the synchronous approach with TAU~\cite{shende2006tau}.
We find that ADIOS2 lossless compression takes nearly the same amount of time in all cases, while the time to write out compressed data decreases as the maximum allowed error increases. This is expected as the total compression ratio rises, requiring us to write out less data via the IO subsystem.



For the analysis of the hybrid data compression we evaluate the execution time on 24 nodes when 1, 9, 18, and 36 core(s) out of the 72 cores on each of the used nodes on Raven supercomputer are allocated for the asynchronous part of the data compression, i.e., the lossless compression done by ADIOS2. Because there are fewer cores for simulation, the execution time increases with the number of cores assigned to the in-situ data compression, as shown in the right graph of Fig.~\ref{compre_res}.  
Furthermore, 
similar to the synchronous in-situ data compression, the execution of the simulation consumes the majority of the time. However, even the best hybrid approach takes longer than the synchronous approach, as additional MPI communication is required.



\begin{figure*}[t]
	\includegraphics[width=0.98\textwidth]{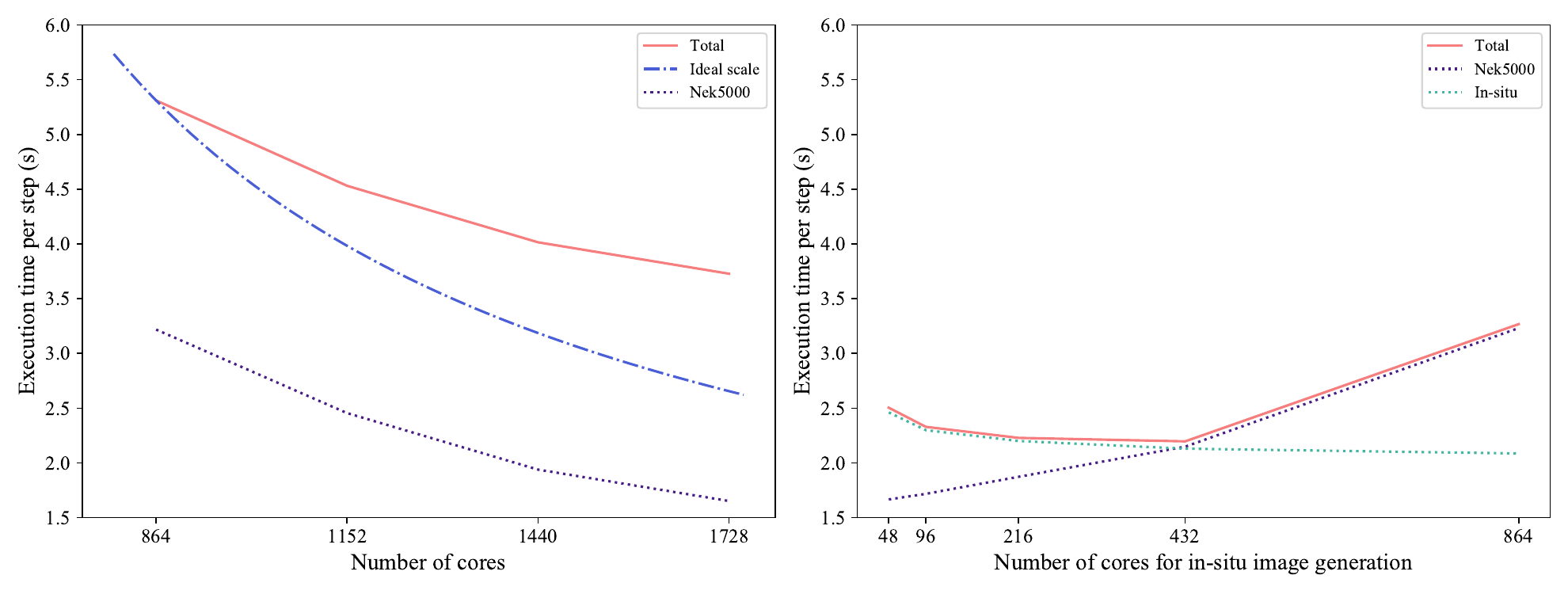}
	\vspace{-1.5em}
	\caption{Execution time of Nek5000 with synchronous in-situ image generation every two steps on Raven supercomputer (left) and asynchronous in-situ image generation every two simulation steps on 24 Raven nodes (right).}
	\label{visual_res}
	\vspace{-1.5em}
\end{figure*}

\begin{figure}[bp!]
    \vspace{-2.em}
	\includegraphics[width=0.48\textwidth]{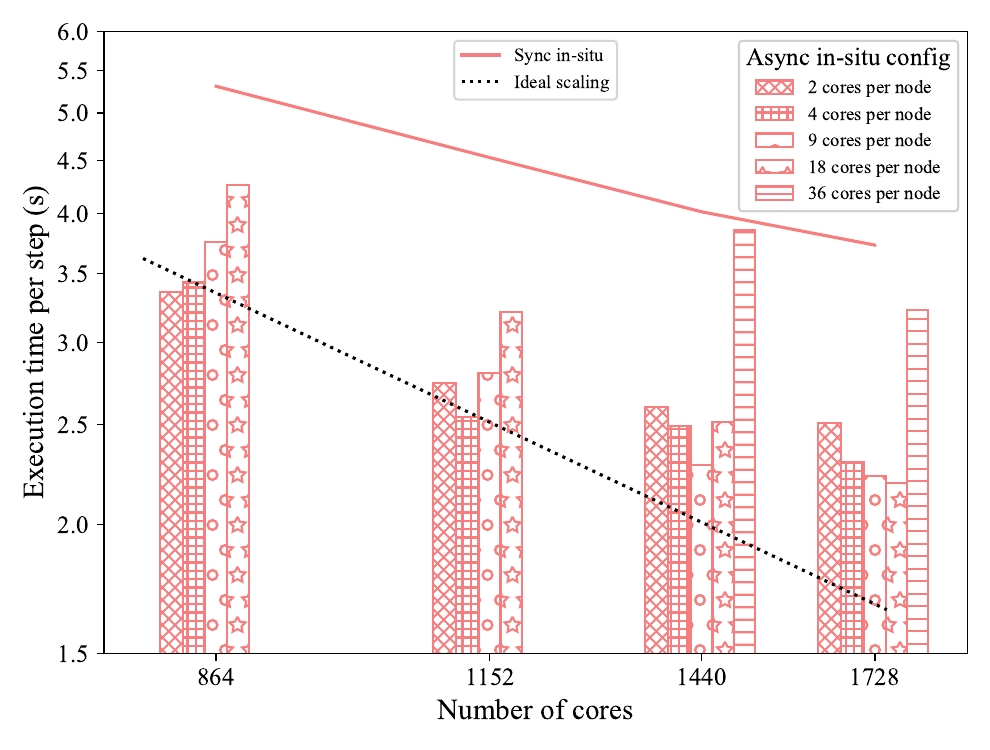}
	\vspace{-1.5em}
	\caption{Logarithmic execution time of Nek5000 with synchronous and asynchronous in-situ image generation every two steps on Raven supercomputer. (Due to the memory limitation, the test cannot be done on 12 and 16 nodes with 36 cores per node for asynchronous in-situ.)}
	\vspace{-2.em}
	\label{visual_res2}
\end{figure}

\subsection{In-situ image generation}

Then we study synchronous and asynchronous in-situ image generation for turbulent fluid in a bent pipe simulation with Nek5000.

\paragraph{Implementation}
In the \textit{synchronous image generation case}, we used the in-situ adaptor functions from the in-situ package repository developed by \emph{Atzori et al.} \cite{atzori2022situ} as the \textsf{Nek-proc} adaptor functions in Fig.~\ref{sync} to connect the Fortran-based Nek5000 and the C/C++-based in-situ task. Because of the different grid used by Nek5000, the pressure scalar and velocity vector fields are mapped into the VTK grid, with a deep copy of the simulation results in the adaptor functions.

In the \textit{asynchronous image generation case}, we construct two groups of adaptor functions for the simulation solver and in-situ task. The \textsf{Nek-writer} adaptor functions shown in Fig.~\ref{async} are a group of Fortran and C/C++ functions. They connect the simulation solver and the writer in the ADIOS2 writer-reader pair and pass the pressure and velocity data using the Nek5000 data structure. C/C++ only \textsf{Reader-proc} adaptor functions connect the reader in the ADIOS2 writer-reader pair and the image generator. 
The VTK unstructured grid is generated during image generator initialization based on the number of elements in one core dedicated to the image generator. The image generator's adaptor functions also perform a deep copy to convert the fields to VTK format. 

\paragraph{Evaluation} 
We ran the same strong scalability test on Raven with 12, 16, 20, and 24 nodes as before. 
The left graph in Fig.~\ref{visual_res} depicts the performance of the simulation with synchronous in-situ image generation every two steps. Although the Nek5000 scales well, the execution time to generate images with ParaView/Catalyst does not scale and remains nearly constant. The MPI collective communication was identified as the bottleneck in the previous study~\cite{atzori2022situ}. This also corresponds to our poorly scaling overall execution time for image generation as the number of cores increased.

We also evaluated the execution time, when 2, 4, 9, 18, and 36 cores in 72 cores on each of 24 Raven nodes (i.e., 48, 96, 216, 432 and 864 cores) are used for the asynchronous image generation every two simulation steps. To better understand the performance of the simulation with asynchronous image generation, we measured total execution time, simulation Nek5000 time, and in-situ image generation time.

The right graph in Fig.~\ref{visual_res} shows that the time to generate images every two simulation steps scales poorly with the number of cores, while simulation efficiency decreases as the number of cores devoted to image generation increases (and thus not to the simulation). The total execution time is the maximum time of simulation and image generation. Thus, as the number of cores for in-situ image generation increases, the total execution time decreases until it no longer scales and the negative effect on simulation time takes precedence. The total execution time is minimal, with one quarter of cores on 24 nodes for in-situ image generation, and the asynchronous image generation and simulation take the same amount of time. 

To study the configuration with the best performance when the total number of cores are changed, we repeated the configuration evaluation with 12, 16, 20 and 24 Raven nodes. 

Fig.~\ref{visual_res2} compares the total execution time of synchronous image generation with total execution time of asynchronous image generation every two simulation steps. 
The asynchronous in-situ approach outperforms the synchronous approach. 
The best performances of 12, 16, 20 and 24 nodes appear with 2, 4, 9 and 18 cores on each node for in-situ image generation respectively. 
The best total execution times of simulation with asynchronous approach are approximately 60\% shorter than the synchronous approach, and scalability is improved, but it cannot scale ideally due to the communication cost of the MPI collective communication. 

\begin{figure}[bp!]
    \vspace{-2.em}
	\includegraphics[width=0.48\textwidth]{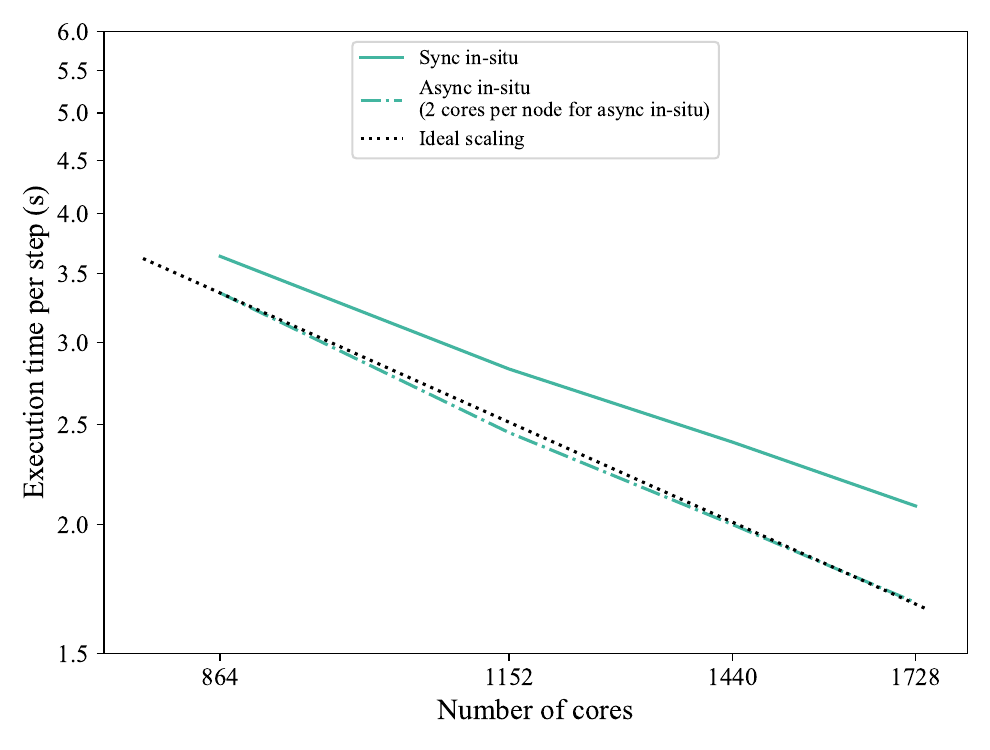}
	\vspace{-1.5em}
	\caption{Logarithmic execution time of Nek5000 with synchronous and asynchronous in-situ image generation every five steps on Raven supercomputer. (The configuration of the asynchronous in-situ approach is two cores per node for in-situ task.)}
	\vspace{-2.em}
	\label{visual_res3}
\end{figure}

We repeated the whole experiment sets with image generation every five simulation to investigate the influence of the in-situ task frequency. As shown in Fig~\ref{visual_res3}, the asynchronous in-situ approach also outperforms the synchronous approach with this frequency. The best performances of 12, 16, 20 and 24 nodes all appear, when two cores on each node are used to generate the image. 
The simulation with asynchronous in-situ every five simulation steps has a lower in-situ workload and MPI communication cost than every two simulation steps, so it has strong scalability with the number of nodes we used. 


\subsection{In-situ uncertainty quantification} 
We also study synchronous, asynchronous and hybrid in-situ uncertainty quantification for a turbulent fluid in a bent pipe simulation with Nek5000.

\paragraph{Implementation} 
In the \textit{synchronous UQ}, the \textsf{Nek-proc} adaptor functions shown in Fig.~\ref{sync} pass the data from Fortran to C/C++ and use C/C++ functions as bridge function to embed Python, since the UQ analyzer is programmed in Python. The simulation results are passed as a single-dimensional Numpy array to the data analyzer to update the training lags.

Two groups of adaptor functions are used in the \textit{asynchronous UQ} for the simulation solver and data processor. The \textsf{Nek-writer} adaptor functions in Fig.~\ref{async} attached to the simulation solver and the resulting workflow are similar to the mixed Fortran and C/C++ adaptor functions in the asynchronous image generator. The \textsf{Reader-proc} adaptor functions connected to the UQ data analyzer are programmed in Python. To simplify the workflow, the ADIOS2 Python APIs are used to build the reader in the ADIOS2 writer-reader pair. 

\begin{figure*}[t]
	\centering
	\includegraphics[width=0.98\textwidth]{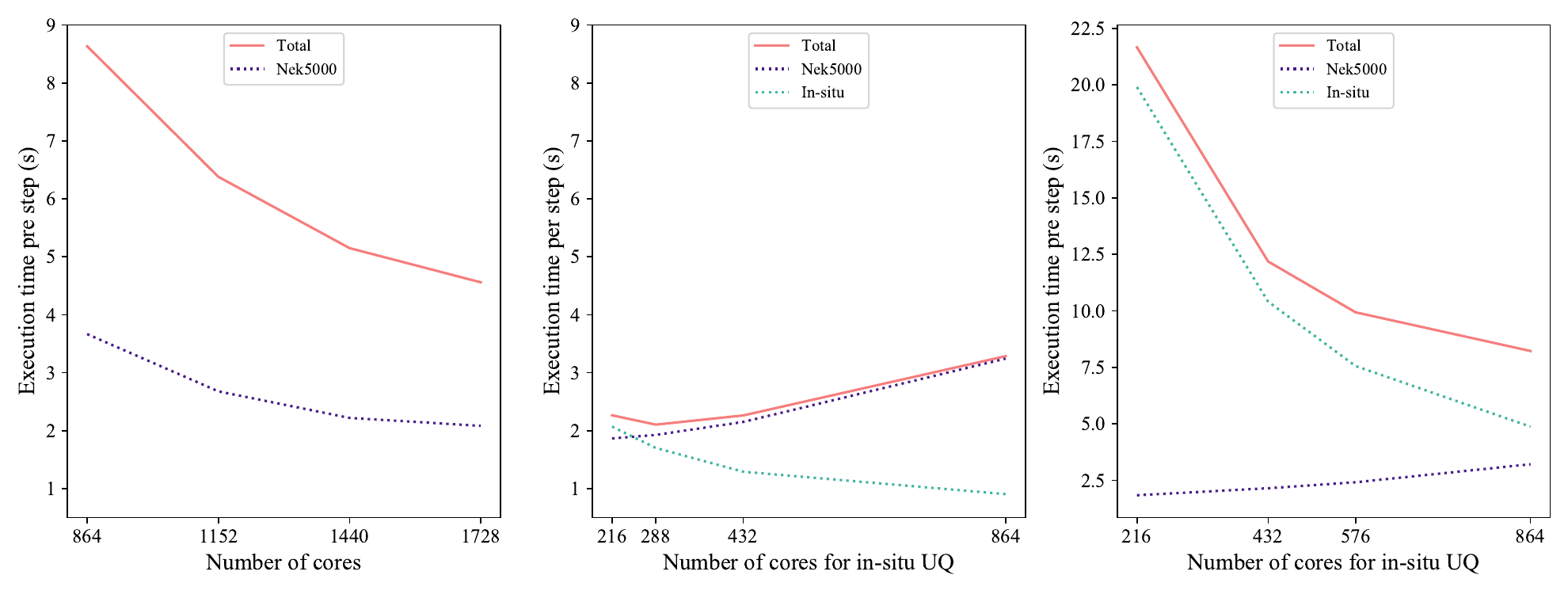}
	\vspace{-1.5em}
	\caption{Execution time of Nek5000 with synchronous in-situ uncertainty quantification (left), asynchronous in-situ uncertainty quantification on 24 Raven nodes (middle) and hybrid in-situ uncertainty quantification on 24 Raven nodes (right).}
	\vspace{-1.em}
	\label{UQ_res}
\end{figure*}

In the \textit{hybrid UQ}, three groups of adaptor functions shown in Fig.~\ref{hybrid} allow for synchronous execution of the training lag updating and the asynchronous execution of the model estimation and uncertainty calculation from the model in the UQ analysis. The \textsf{Nek-proc} adaptor functions are similar to the adaptor functions in synchronous UQ. To connect Fortran and Python functions, they use C/C++. The \textsf{Proc-writer} and \textsf{Reader-proc} adaptor functions are programmed in Python. The \textsf{Proc-writer} adaptor functions pass the training lags to the Python-based ADIOS2 writer; the \textsf{Reader-proc} adaptor functions transfer the training lags from the Python-based ADIOS2 reader to the asynchronous part of UQ analysis.


Every simulation step, we update one training lag from the average velocity of each element and then quantify the uncertainty from 50 training lags. This frequency is rather high for UQ. We used it as a stress test. We also examined the performance with a standard UQ frequency (i.e., updating one training lag every 20 simulation steps and estimating the uncertainty from 25 training lags every 500 simulation steps). 

\paragraph{Evaluation} 

We investigated the scalability of the simulation with synchronous in-situ UQ on 12, 16, 20, and 24 nodes. 
The total execution time of the stress test and the execution time of the simulation scale well, as shown in the left graph in Fig.~\ref{UQ_res}. But UQ takes longer than simulation, and in the profiling reports, the execution time of UQ on each core varies. Because the estimation portion of UQ includes model estimations involving regression, the workload to calculate the uncertainty is unknown. As a result, the load balancing is dependent on simulation results and is frequently suboptimal. 

We used 9, 18, 24, and 36 cores in 72 cores on each of 24 Raven nodes for asynchronous UQ with stress test frequency.
The right graph in Fig.~\ref{UQ_res} shows that the performance of the asynchronous UQ is even worse than the synchronous approach. 
The in-situ UQ takes longer than the Nek5000 simulation in this approach.
Although its performance improves as the number of cores for in-situ increases, the total execution time with 36 cores for asynchronous in-situ is still twice as long as the synchronous approach. This is due to the large workload difference between the UQ steps. To ensure data consistency, the simulation cores need to communicate with the in-situ UQ core at each simulation step. This keeps the simulation and UQ from running concurrently. 

\begin{figure}[bp!]
\vspace{-2.0em}
	\includegraphics[width=0.48\textwidth]{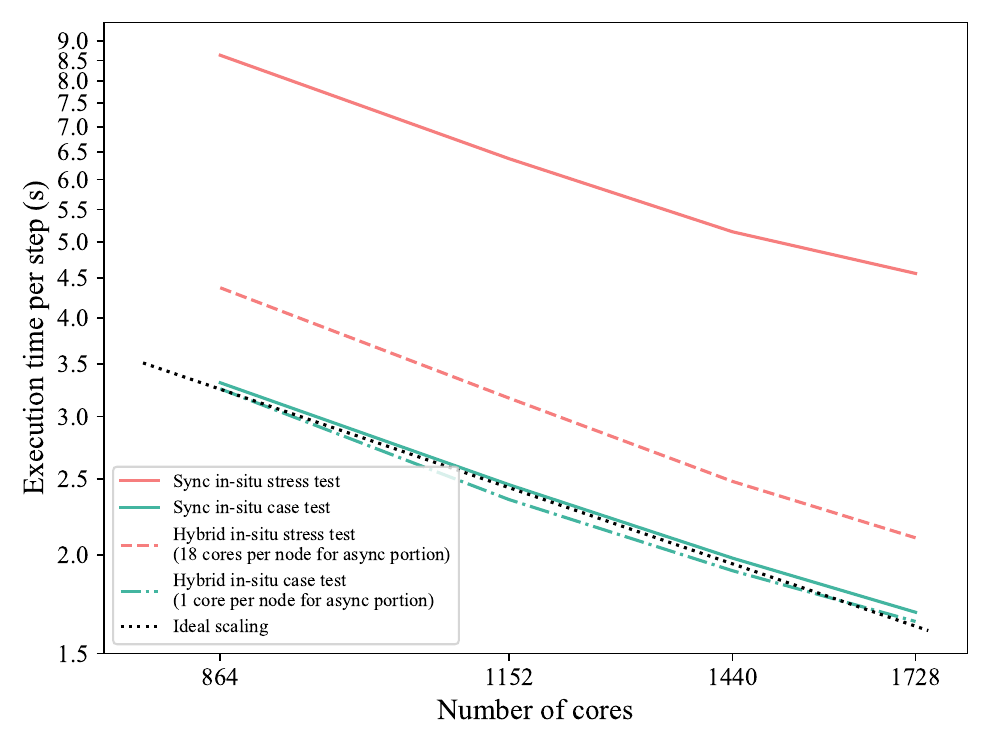}
	\vspace{-1.5em}
	\caption{Logarithmic execution time of Nek5000 with synchronous and hybrid in-situ uncertainty quantification.}
	\label{UQ_res2}
\end{figure}

We used 9, 12, 18, and 36 cores within 72 cores on each of 24 Raven nodes for a hybrid setup, where updating the training lags is done synchronously and model estimation and uncertainty calculation is done asynchronously. 
By lowering the communication frequency, the hybrid approach enables concurrency. The communication between simulation and in-situ cores is only required before the asynchronous section of the UQ. The middle graph in Fig.~\ref{UQ_res} shows that the total execution time on 24 nodes decreases with the number of cores for the asynchronous portion until the simulation takes almost the same amount of time as the asynchronous UQ portion; then the total execution time increases because the simulation time takes longer in this phase and increases with the ratio of cores for in-situ tasks. When compared to the synchronous approach, the hybrid approach improves the cache hit ratio, and better balances the workload. The data transfer in the hybrid in-situ approach is also optimized when compared to the asynchronous approach. The total amount of data required to be transferred in the hybrid approach is significantly less than in the asynchronous approach, and in comparison to the asynchronous approach's frequent small trunk data transfer, the infrequent larger trunk data transfer in the hybrid approach results in lower latency.

We repeated the evaluation with 12, 16, 20 and 24 Raven nodes, and performed the UQ with both stress test frequency and common frequency, to study the influence of the total number of cores and in-situ task problem size on the configuration with the best performance. Because the common frequency leads to a relatively cheaper computational cost, we add one core per Raven node as additional configuration in the tests.

Fig.~\ref{UQ_res2} compares the total execution time of synchronous with the best total execution of hybrid UQ of both tests. The hybrid in-situ UQ outperforms the synchronous approach. In the case presented here, the best performance of the common case appears with one core on each node for in-situ; the best performance of the stress test appears with twelve cores on each node. The total execution time of simulation with hybrid UQ are approximately 50\% shorter than the synchronous approach in the stress test, and scalability is improved. In the common case test, the hybrid approach still outperforms the synchronous approach slightly.


\section{Discussion and Conclusions}
\label{sec:conclude}

In this paper we focus on the resource distribution for the synchornous, asynchronous and hybrid in-situ approaches on homogenenous, multicore-CPU based HPC systems. 

We can conclude from our in-situ data compression that synchronous execution is favorable for comparably small in-situ tasks for the sake of the performance because in asynchronous or hybrid executions, not only additional communication overhead is introduced, but the number of resources available for the simulation is also reduced, while the dedicated resources for the in-situ task are underutilized. 

From our in-situ image generation, we can conclude that for larger in-situ tasks that do not scale well, an asynchronous approach is preferable to a synchronous one because fewer resources can be assigned to the in-situ task, to limit the effects of poor scalability. The sweet-spot for how many resources are assigned to the in-situ task (and thus remain for the simulation) must be determined, and the sweet-spot distribution might change with the total number of resources due to the different scalability of the simulation and in-situ tasks. 

We can conclude from our in-situ data compression and uncertainty quantification that hybrid execution is the preferred model for cases where the implementation of the in-situ tasks are strongly dependent on functions from the main solver or are performed frequently, but other parts can overlap the execution of the simulation and benefit in the performance from the asynchronous execution. When the scalability of the simulation and in-situ tasks is similar, the sweet-spot resource distribution is stable. With this property, when larger number of resources are used, the ideal resource distribution could be predicted from the performance of fewer resources without constructing and analyzing the complex performance models.

In general, we explored synchronous, asynchronous, and hybrid in-situ approaches and compared three use cases with different characteristics in this paper. First, we reduced the amount of data and corresponding IO time by using in-situ lossy and lossless compression. Then, we performed in-situ visualization, and finally, we used uncertainty quantification in an in-situ manner. For each of these use cases, we analyzed the benefits of the three in-situ approaches. Due to the comparably lower workload and good scaling behavior, compression performs best in synchronous mode; asynchronous or hybrid approaches just add overhead without significant benefits. For visualization, the asynchronous approach performed best, as it allows optimizing the computing resource allocation to minimize the overhead from the MPI collective communication. For uncertainty quantification, the synchronous approach outperforms the asynchronous one. As the frequencies and computational costs of the two sections of the UQ differ, the simulation and data analysis are not executed concurrently in the asynchronous approach. However, UQ consists of two portions that can be split and thus performed in a hybrid in-situ mode, resulting in  lower total data amount transferred, lower latency from the larger trunk size of data transferred in one communication, and better data access pattern in simulation and data analysis, and thus the best performance. We can thus conclude from these case studies that in-situ tasks with high frequency, low computational cost, or/and low communication overhead may perform better in synchronous approach, whereas  in-situ tasks with low frequency, high computational cost, high communication overhead, or/and low complexity to decouple from simulation could benefit from the asynchronous approach. 

In future work we plan to derive models from our experimental findings that will help to choose among the synchronous, asynchronous, and hybrid in-situ approaches. Furthermore, we investigate in-situ approaches on hybrid computational nodes consisting of GPUs and CPUs. NekRS~\cite{fischer2021nekrs} and NEKO~\cite{jansson2021neko} are promising simulation solvers with GPU support. Current approaches for scientific simulations often only use the GPUs for the simulation, leaving the CPUs underutilized and are thus a perfect target for in-situ tasks. Moreover, we also plan to extend our study of the in-situ techniques on exascale simulations with billions of data points (or equivalent millions of elements in Nek5000) and verify the possibility to predict optimal resource distribution for the exascale case with the performance data from the large cases (such as the ones in this paper). Finally, we plan to continue working with ADIOS2 towards a generic in-situ framework.

\section*{Acknowledgment}
Dr. Saleh Rezaeiravesh and Christian Gscheidle are gratefully acknowledged for providing the UQ case. This work is partially funded by the “Adaptive multi-tier intelligent data manager for Exascale (ADMIRE)” project, which is funded by the European Union's Horizon 2020 JTI-EuroHPC research and innovation program under grant Agreement number: 956748.
The authors would like to express their gratitude the Max Planck Computing and Data Facility (MPCDF) for providing compute time on the Raven Supercomputer. Furthermore, part of the computations were enabled by resources
provided by the Swedish National Infrastructure for Computing (SNIC),
partially funded by the Swedish Research Council through grant agreement
no. 2018-05973.

\bibliographystyle{IEEEtranS}
\bibliography{refs}

\vspace{12pt}
\color{red}

\end{document}